\documentclass[twocolumn,showpacs,preprintnumbers,amsmath,amssymb,prb]{revtex4}
\usepackage{graphicx}
\usepackage{dcolumn}
\usepackage{bm}
\begin{document}
\title{Optical model-solution to the competition between a pseudogap
 phase and a Mott-gap phase in high-temperature cuprate
 superconductors}
\author{Tanmoy Das, R. S. Markiewicz, and A. Bansil}
\address{Physics Department, Northeastern University, Boston MA
02115, USA}
\date{\today}

\begin{abstract}

We present a theoretical framework for a quantitative
understanding of the full doping dependence of the optical spectra
of the cuprates. In accord with experimental observations, the
computed spectra show how the high-energy Mott features continue
to persist in the overdoped regime even after the mid-infrared
(MIR) peak originating from the pseudogap has collapsed in a
quantum critical point. In this way, we reconcile the opposing
tendencies of the MIR and Mott peaks to shift in opposite
directions in the optical spectra with increasing doping. The
competition between the pseudogap and the Mott gap also results in
rapid loss of spectral weight in the high energy region with
doping.

\end{abstract}

 \maketitle\narrowtext

\maketitle \narrowtext

\section{Introduction}

Most experiments on the cuprates show that when the Mott
insulating state is doped with electrons or holes the gap feature
in the spectrum moves to lower energies and collapses at a quantum
critical point, consistent with the notion that electron
correlation effects in the Mott state weaken systematically with
doping.\cite{zxshen,nparm,tacon} In sharp contrast, optical
experiments reveal a substantially more complex picture in that
the doping of the Mott insulator induces a new mid-infrared (MIR)
feature\cite{haule,chakraborty,mishchenko,vidmar}, which collapses
with doping much as seen in other experiments, but at the same
time a high-energy Mott feature persists even into the overdoped
system with its spectral weight shifting to the MIR feature with
increasing doping.\cite{cooper,hwang,arima,uchida,onose,onoseprb}
In fact, the Mott feature not only persists in the optical
experiments, but it moves to {\it higher} energies with doping,
suggesting that strong electron correlation effects continue to
play an important role in the cuprates at all dopings. Modelling
and understanding the optical spectra of the cuprates thus becomes
of key importance in unravelling the routes by which the Mott
insulator turns itself into a superconductor, a transition that
remains poorly understood.

The question of correlations has been traditionally framed in
terms of a Slater picture of itinerant electrons where an
insulator forms a gap via the development of a long range magnetic
order, or a Mott picture in which the metal-insulator transition
is driven by a local condition of no double occupancy. Although
the Slater or the Mott picture is often invoked, many materials
lie in the crossover regime where the electronic states are not
well described as being either fully itinerant or fully localized.
Such materials are often endowed with unique and exotic properties
and are of great current interest from the viewpoint of
fundamental physics as well as potential for applications. The
strength of electron correlations in the cuprates has been a
matter of considerable debate since the discovery of these
fascinating materials.\cite{comanac}

We have obtained the optical spectra within the framework of a one
band Hubbard model where the self-energy is obtained
self-consistently in an intermediate coupling scenario to account
for spin and charge fluctuations via a computation of the
susceptibility over the entire doping range. We emphasize that the
present model essentially does not involve any free
parametrization in that the bare LDA dispersion is taken to be the
CuO$_2$ band (in the tight-binding form) and a doping-independent
bare $U$ is chosen to reproduce the experimental charge-transfer
gap at half-filling. The screened $U$ at various doping levels is
then computed due to charge
fluctuations\cite{kanamori,markiecharge}, which self-consistently
induce a low-energy magnetic order in the in-gap states resulting
in a pseudogap in the underdoped system. Our computations
reproduce quantitatively the experimentally observed doping
evolution of not only the shifts in the positions of the MIR and
Mott features, but also the rapid transfer of spectral weight from
the high to the low energy region with increasing doping. The
theoretical shifts of the MIR feature with doping are in accord
with angle-resolved photoemission spectroscopy (ARPES)\cite{waterfall} and other
spectroscopic probes. We emphasize that existing approaches
invoking either strong or weak coupling scenarios fail to capture
the essence of the doping evolution of the optical spectra.

\section{Optical spectra in electron and hole doped cuprates}

Fig.~1 shows one of our key results. The computed evolution of the
optical conductivity $\sigma(\omega)$ is seen to be in excellent
accord with measurements on electron-doped Nd$_{2-x}$Ce$_x$CuO$_4$
(NCCO) and hole doped La$_{2-x}$Sr$_x$CuO$_4$ (LSCO)
\cite{onoseprb,uchida}. All the spectra show a nearly isosbetic or
equal absorption point near 1.3 eV (1~eV for LSCO) [black vertical
line], consistent with the experimental behavior\cite{foot2}. The
doping evolution is completely different on opposite sides of this
isosbetic point. Above this point, the spectrum is dominated by a
broad hump feature associated with the Mott gap. At half-filling,
only this feature is present and the calculated optical spectra
show an insulating gap whose energy, structure, and intensity
match remarkably with measurements\cite{onoseprb,uchida}. As
doping increases, the high energy peak shifts to higher energy and
broadens and its spectral weight is transferred to the Drude and
MIR peaks. The MIR peak shifts to lower energy with doping and
gradually sharpens. Note that in both samples at the highest
doping, when the MIR peak collapses into the Drude peak, Mott-gap
features still persist in the spectrum. The doping evolution also
shows similar behavior in other
cuprates\cite{cooper,hwang,arima,mcguire,uchida,onose,onoseprb}.
Notably, our calculations also describe the anomalous $\sigma\sim
1/\omega$-dependence found in most cuprates associated with
magnetic scattering.

\begin{figure}
\rotatebox{270}{\scalebox{0.5}{\includegraphics{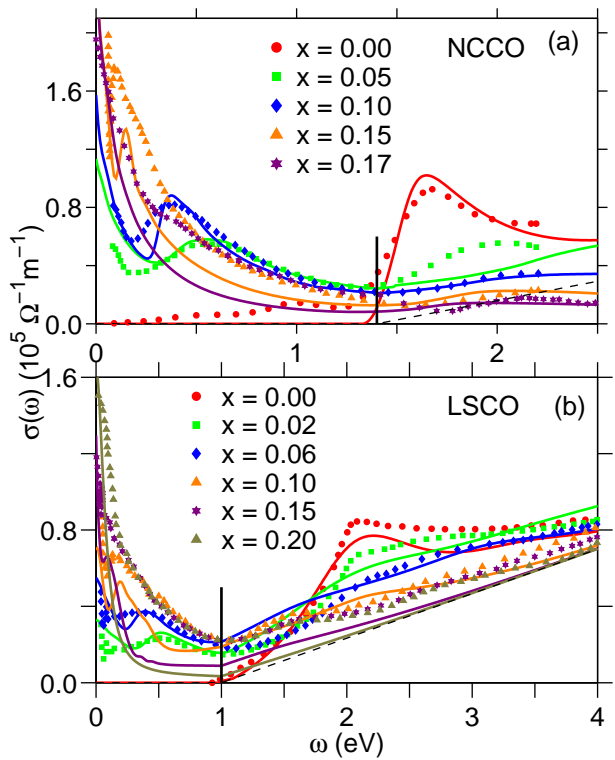}}}
\caption{(color online) Comparison between calculated optical
spectra and experiments in NCCO in (a) and LSCO in (b).
Experimental results for NCCO are taken from
Ref.~\onlinecite{onoseprb}, except for $x=0.17$, which including
LSCO data, are from Ref.~\onlinecite{uchida}. The $x=0.17$ dataset
for NCCO includes a background subtraction to match the former
dataset\cite{chen}. Dashed line gives the background contribution
added to the theoretical spectrum at $x=0$
[Ref.~\protect\onlinecite{comanac}].} \label{opsp} \end{figure}

\subsection{Origin of optical dichotomy}

Fig.~2 helps delineate the microscopic origin of these features by
comparing the spectral intensities relevant for ARPES and optical
spectra at a representative doping of $x=0.10$ for NCCO; similar
analysis for LSCO is not shown for brevity. In the computed ARPES
spectrum in Fig.~2(a), the underlying LDA dispersion is clearly
visible, but the spectral weight has split into four subbands, as
is also the case in variational cluster calculations\cite{grober}.
The highest and lowest bands are an incoherent residue of the
undressed bands, which we will refer to as upper and lower Hubbard
bands. The two inner bands are coherent in-gap states split by a
spin density wave induced AFM gap into upper and lower magnetic
bands. The in-gap states and the high energy Hubbard bands are
separated by high-energy kinks (`waterfalls') predominantly
associated with magnetic excitations, as observed universally in
all cuprates by ARPES\cite{graf,moritz,markiewater}, and found in
quantum monte carlo (QMC)\cite{macridin} and
variational\cite{grober} calculations.

\begin{figure*}
\rotatebox{0}{\scalebox{0.9}{\includegraphics{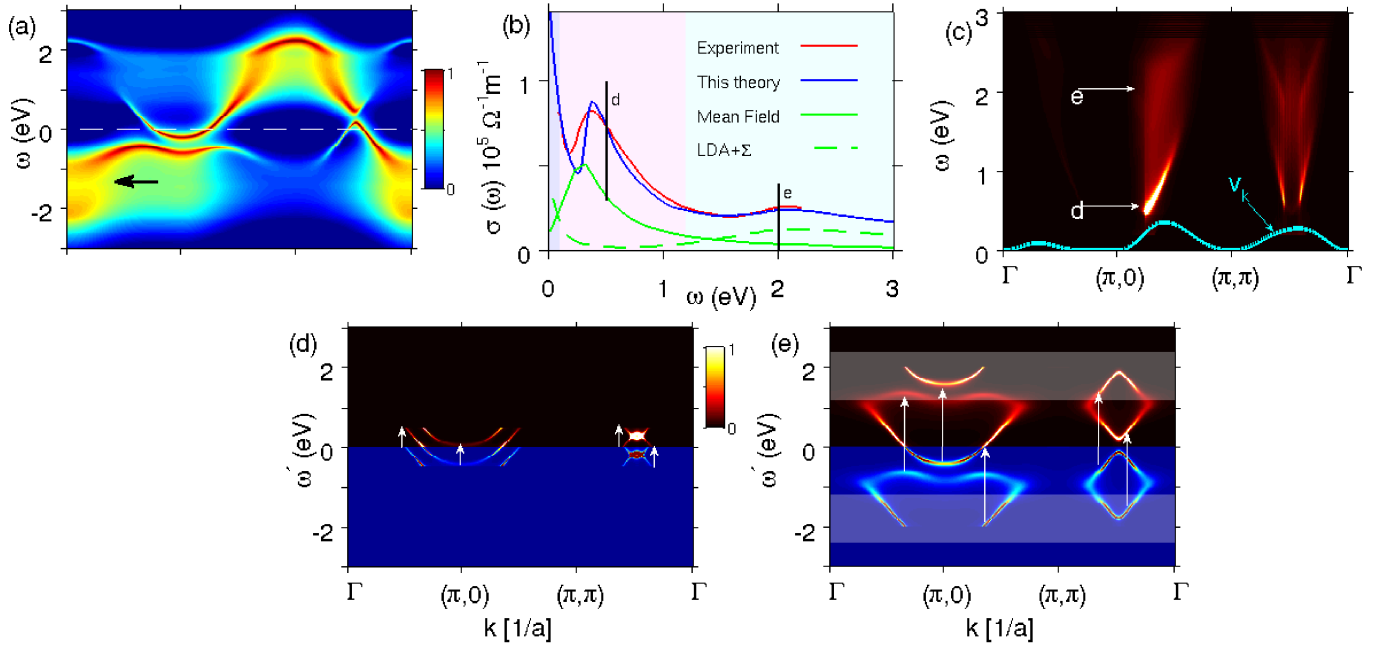}}}
\caption{(Color online) Connection between optical spectra and
spectral intensity maps in NCCO. (a) Computed spectral intensity
(on log scale) as a function of $\omega$ along the high symmetry
lines at a representative doping of $x=0.10$. (b) Calculated
optical spectra are compared with the corresponding experimental
results\cite{onoseprb}. Green solid line gives the spectrum
calculated with coherent bands only, i.e. without the self-energy.
Dashed green line is based on using LDA bands with self-energy
correction but without the pseudogap. The shaded regions of
different colors (left to right) approximately mark Drude, MIR and
high energy regions. Two vertical black lines indicate the photon
energies at which frames (d) and (e) are calculated. (d), (e) give
source and sink maps corresponding to the optical transitions at
fixed photon energy $\omega$ along the high symmetry directions.
White vertical arrows indicate the photon energy connecting source
and sink points involved in a particular transition. White shaded
region in (e) highlights the incoherent part of the spectral
weight, not visible in (d). (c) Optical spectrum as a function of
photon energy is plotted along the high symmetry lines. Cyan line
gives the band velocity as a function of $k$ (arbitrary units).
The two white arrows indicate the two photon energies at which (d)
and (e) are calculated. } \label{arpesop}
\end{figure*}
The corresponding optical spectrum in Fig.~2(b) consists of three
main regions marked by shading of different color: (1) The low
frequency Drude region for $\omega \lesssim 30$meV; (2) The MIR
region; and, (3) the high energy Mott gap region for $\omega
\gtrsim 1.5$eV. We concentrate here on the MIR and Mott-gap
regions, and return below to comment on the Drude region. The
interband optical absorption is proportional to the joint density
of states (JDOS), so that at each energy we can construct a
`source map' showing the filled states which make a strong
contribution to the transition, and a `sink map' showing the
contribution of the corresponding empty states. Figs.~2(d) and (e)
show the states responsible for the optical transitions along the
high symmetry lines at representative photon energies of
$\omega=0.5$~eV near MIR peak and $\omega=2$~eV around the high
energy peak (vertical bars in Fig.~2(b)). At $\omega=0.5$~eV, the
transitions are confined within the in-gap states only, whereas at
$\omega=2$~eV, the optical subbands involve predominantly the
incoherent region. The depletion in the optical spectral weight
near the isosbetic point in Fig.~1 and in Fig.~2(b) is thus
associated with the `waterfall' region marked by arrows in the
ARPES spectrum in Fig.~2(a)\cite{foot2}.

The total optical spectral weight obtained in Fig.~2(b) is the
integral of the JDOS times the band velocity. The role of the
latter factor is explicated in Fig.~2(c), where contributions to
$\sigma$ are plotted as a function of photon energy $\omega$ and
momentum $k$. Although the source-sink map shows a symmetry about
the $(\pi,0)$ point, the quasiparticle velocity is low along the
$\Gamma\rightarrow(\pi, 0)$ direction [cyan solid line in
Fig.~2(c)], leading to greatly reduced spectral intensity
associated with those regions. In contrast, the large
quasiparticle velocity in the other two directions is responsible
for two distinctive intense streaks (labelled d and e). The MIR
peak is clearly dominated by the antinodal quasiparticles, whereas
the high energy hump stems from a wider $k$-range.

\subsection{Competition between Pseudogap amd Mott gap}
\begin{figure}
\rotatebox{0}{\scalebox{0.45}{\includegraphics{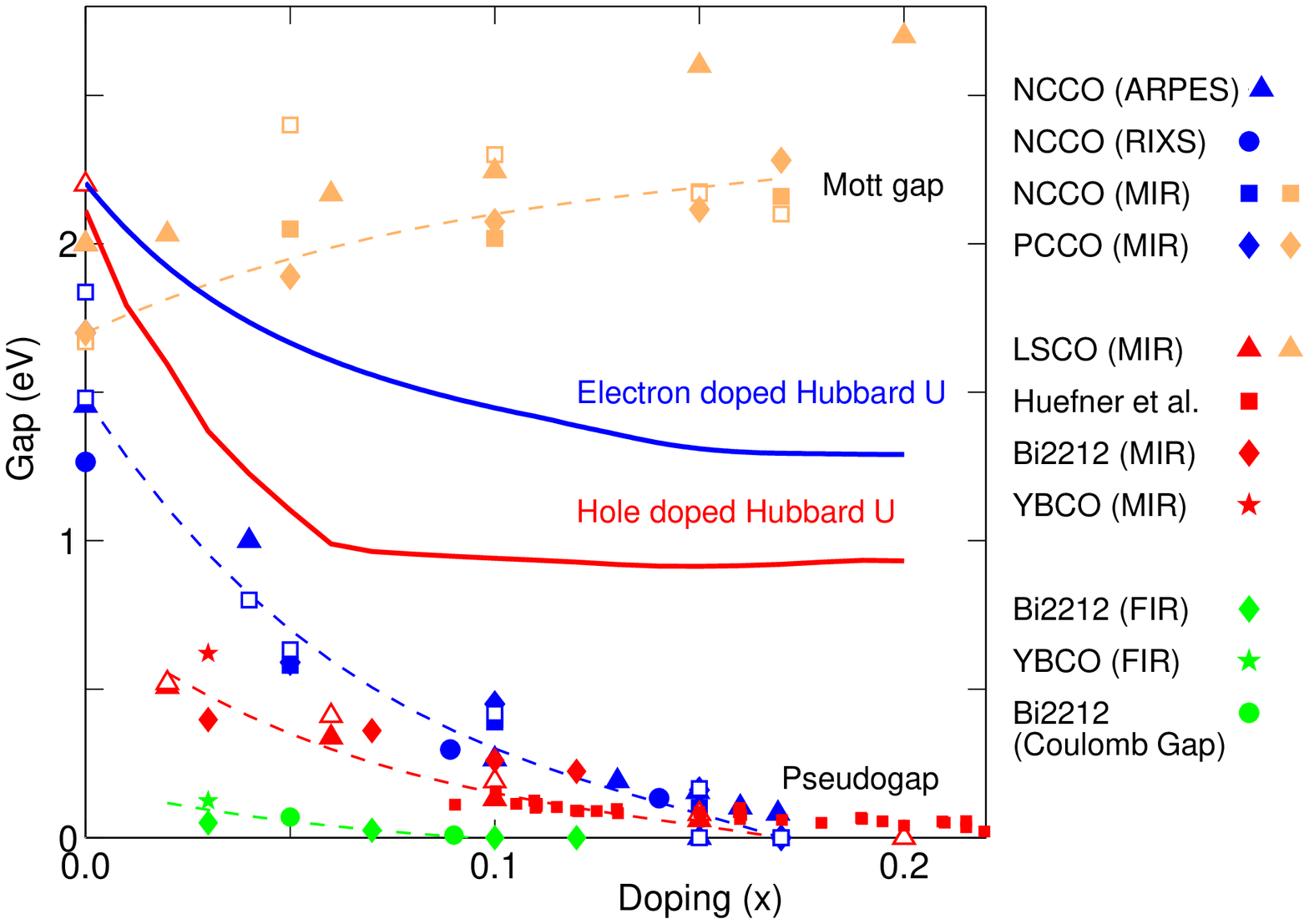}}}
\caption{(Color online) Contrasting doping dependences of the MIR and Mott gaps in the cuprates. Present theoretical results  are compared to optical data\cite{arima,uchida,onoseprb} and pseudogap data measured by 
ARPES\cite{nparm,matsui} and RIXS\cite{RIXS} for electron doped NCCO and many other probes for hole doped cuprates. The red squares are the pseudogap for four different layered compounds [Bi2212, YBCO, TBCO and HBCO] as they are obtained from ARPES, tunneling, Raman, Andreev reflection, and heat capacity experimental data (reproduced from a detailed survey, Ref.~\onlinecite{huefner}). The red diamond and star symbols represent independent MIR and  green diamond and star give FIR gaps measured simultaneosly in optical spectra of Bi2212 and YBCO\cite{lupi}, while the green circles represent a Coulomb gap seen in ARPES\cite{ding} [energy values are obtained assuming a ratio $2\Delta/k_BT=4$].  In all cases, open symbols of same colors are the corresponding theory which are obtained self-consistently at each doping.  Shown also are the computed
screened value of $U$ (solid lines)\cite{markiecharge,kanamori} used in these calculations. The dashed lines are guides to the eye for the Mott gap (gold), the pseudogaps for electron (blue) and hole (red) doping, and the FIR and Coulomb gaps (green). }
\label{gap}
\end{figure}
Fig.~3 summarizes our results and elaborates upon the competition between the collapse of
the pseudogap in the coherent bands at a critical doping, and the
persistence of the Mott gap in the incoherent bands. The gaps
extracted from the optical spectra are now compared with other
experimental data for various electron- and hole-doped cuprates. The
pseudogap seen in the optical spectrum corresponds to the true AFM
gap, and its doping evolution is in good agreement with optical,
ARPES\cite{nparm,kusko,matsui} and resonant inelastic x-ray
scattering (RIXS)\cite{RIXS} results. Both ARPES and MIR data
predict a QCP near $x=0.17$\cite{tanmoyprl,tanmoytwogap}
consistent with the present theoretical predictions. In contrast,
the Mott gap shows the opposite doping dependence to the
pseudogap, increasing slowly with doping. Although the Mott gap
does not show a real QCP, it rapidly loses intensity with doping.

Similar results are found in hole doped LSCO\cite{uchida}, and other layered cuprates, including
Bi$_2$Sr$_2$CaCu$_2$O$_{8+\delta}$ (Bi2212)\cite{hwang}, YBa$_2$Cu$_3$O$_{6+x}$ (YBCO)\cite{cooper},
Tl$_2$Ba$_2$CuO$_{6+\delta}$ (TBCO), and HgBa$_2$Ca$_2$Cu$_3$O$_{8+\delta}$ (HBCO)\cite{mcguire} as well as
in x-ray absorption spectroscopy\cite{chen} and QMC computations.\cite{jarrell} Figure~3 also includes data on the MIR gap for LSCO, Bi2212 and YBCO and the pseudogap obtained from a wide range of experimental data, including a recent detailed survey\cite{huefner} (see Fig.~3 caption).  In the intermediate and high doping regimes, there is very good agreement between the MIR and other measures of the pseudogap, while the MIR data provides important evidence for the rapid growth of the pseudogap in the deeply underdoped regime, consistent with theory.  We note that while all the hole-doped cuprates seem to have a very similar doping dependence of the pseudogap, there are subtle differences with electron doped cuprates, including the steepness of the rise at low doping and the exact position of the QCP.\cite{huefner,tallon}  The former difference
seems to be correlated with the doping dependence of the screened Hubbard $U$'s, also shown in Fig.~3.  Since the doping dependence is mainly a screening effect,\cite{markiecharge,kanamori} the difference is probably due to the strong screening associated with the van-Hove singularity (VHS) on the hole-doped side.

While a strong case can be made that in electron-doped cuprates the pseudogap is associated with a coexisting $(\pi ,\pi )$ AFM order, the nature of the pseudogap in hole doped cuprates is not well understood.  There is growing consensus that
it originates from some form of density-wave like competing order (which could include coupling to
phonons)\cite{lee,muller,jarrell1}, and may actually also involve several competing modes, again due to proximity to the VHS\cite{GARPA}. 
However, in our earlier study in Ref.~\onlinecite{tanmoytwogap}, which compared several possible competing order phases, we demonstrated that the shape and doping dependence of the pseudogap is remarkably insensitive to the particular order so long as the competing order vanishes in a QCP near optimal doping.

While our present calculations reproduce the gap structures of cuprates, particularly on energy scales $\ge 100$~meV, additional effects can arise at lower energies, including coupling to phonons and impurities.  In particular, Lupi {\it et al.}\cite{lupi} find in several cuprates that in addition to the MIR feature there is an additional  far-infrared (FIR) peak which appears near 10\% doping and shifts to higher energies at lower doping (green diamond and star symbols in Fig.~3).  This seems to be a disorder effect, and has a similar doping dependence to the so-called Coulomb gap seen in ARPES (green circles) in the low-temperature region of Bi2212.\cite{ding}

We emphasize that the dichotomy between the pseudogap and Mott gap
and the presence of the QCP are robust features of cuprates as the
computations do not involve any free parametrization except the
bare doping-independent value of $U=1.7$~eV at half-filling. We
have computed the doping dependence of $U$ due to the screening
effects of charge fluctuations\cite{kanamori,markiecharge} and
obtained effective $U$ values are shown Fig.~3.

\section{Optical sum rule}

\begin{figure}[top]
\rotatebox{0}{\scalebox{0.3}{\includegraphics{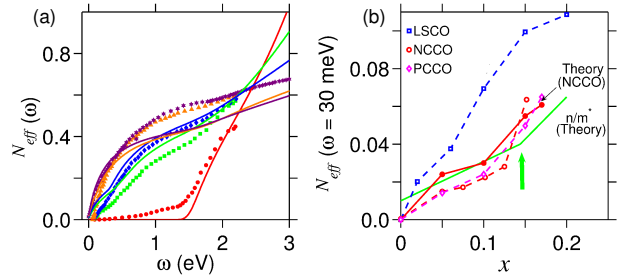}}}
\caption{(Color online) Optical sum rule and spectral weight
transfer. (a) Effective number of electrons $N_{eff}(\omega)$
calculated from the optical spectra in Fig.~1 is compared with
experimental data\cite{onose,onoseprb} on NCCO. Results for LSCO
are similar and are not shown for brevity. (b) Low energy weight
$N_{eff}$ at $\omega=30$ meV as an estimate of Drude weight,
compared with experimental results on various
cuprates\cite{arima,uchida,onose}.  Green line gives the direct
computation of the normalized Drude weight $n/m^*$.} \label{fopsw}
\end{figure}
Finally, we use the integrated optical spectral weight to
illustrate the Mott gap collapse. The effective electron number
(per Cu atom), $N_{eff} (\omega)$, can be defined in terms of the
optical conductivity integrated up to an energy $\omega$:
\begin{equation}\label{eq:3}
N_{eff}(\omega) = \frac{2m_0V}{\pi e^2\hbar N}
\int_0^{\omega}\sigma(\omega^{\prime})d\omega^{\prime},
\end{equation}
where $m_0$ and $e$ are the free electron mass and charge,
respectively, and $N$ is the number of Cu-atoms in a cell of
volume $V$. The results in Fig.~4(a) show how rapidly spectral
weight shifts to low energies with doping, correctly reproducing
the experimental behavior as well as dynamical mean-field
calculations\cite{comanac}, but incompatible with strong coupling
models (such as the $t-J$ or $U\rightarrow\infty$ Hubbard model)
where one assumes no double occupancy of Cu sites. The present
intermediate coupling model of Mott gap collapse on the other hand
properly captures these spectral weight transfers as a function of
doping.

Our model also predicts the Drude weight $\propto\sum_{\bf k}
n_{\bf k}/m^*_{\bf k}$, which can be compared to an experimental
estimate, taken as $N_{eff}$ at a characteristic energy $\omega
=30$~meV in Fig.~4(b) for NCCO\cite{onose}, and with results for
Pr$_{2-x}$Ce$_x$CuO$_4$ (PCCO)\cite{arima} and LSCO\cite{uchida}.
Even here the agreement is quite good. Interestingly, at low
doping the Drude weight increases almost linearly with $x$. Within
our model, this simply reflects scaling with the area of the FS
pockets in the pseudogap state.  At optimal doping, the weight
shows a sharp jump (green arrow) associated with the appearance of
the hole pocket. This topological FS transition is an intrinsic
feature of the model and has been found in NCCO near $x\approx
0.15$ by several experimental probes such as ARPES\cite{nparm},
Hall effect\cite{tanmoysns}, and superconducting penetration
depth\cite{tanmoyprl}. For LSCO (blue symbols),
experiments\cite{uchida} show a similar linear behavior of the
Drude weight, which also corresponds to the doping dependence of
the area of the FS pocket\cite{yoshida}. The peak around
$x\sim0.2$ corresponds to the doping of the Van Hove singularity.

\section{Intermediate Coupling Model}

We evaluate the self-energy $\Sigma$ as a convolution over the
green function $G$ and the interaction $W\sim U^2\chi$ (including
the full spectrum of charge and spin fluctuations within the
random phase approximation (RPA)) as \cite{vignale,waterfall,arpes},
\begin{eqnarray}\label{selfeng}
&&\Sigma({\bf k},\sigma,i\omega_n)=\frac{1}{2}U^2Z \sum_{{\bf
q},\sigma^{\prime}}^ {\prime}\eta_{\sigma,\sigma^{\prime}}
\int_{0}^{\infty}\frac{d\omega_p}{2\pi}\nonumber\\
&& G({\bf k}+{\bf q},\sigma^{\prime},i\omega_n+\omega_p)
\Gamma({\bf k},{\bf q},i\omega_n,\omega_p){\rm Im}[\chi_{\rm
RPA}^{\sigma\sigma^{\prime}}({\bf q},\omega_p)].\nonumber\\
\end{eqnarray}
where $\sigma$ is the spin index and the prime over the $\vec{q}$
summation means that the summation is restricted to the magnetic
Brillouin zone. Here the spin degree of freedom
$\eta_{\sigma,\sigma^{\prime}}$ takes the value of 2 for the
transverse direction and 1 for both longitudinal and charge modes.
In the underdoped region, the pseudogap is modelled by an
antiferromagnetic (AFM) order parameter, resulting in $G$, $\chi$
and $\Sigma$ becoming $2\times2$ tensors\cite{SWZ}. We define a
total self-energy as $\Sigma^{t}=US\tilde{\tau_1}+\Sigma$, where
$\tilde{\tau_1}$ is the Pauli matrix along the $x-$direction and
$US$ is the AFM gap defined below. The self-energy $\Sigma^t$
contains essentially two energy scales: (i) it gives rise to the
SDW with an additional renormalization of the overall
quasiparticle dispersions in the low energy region, and (ii) at
higher energies it produces the Hubbard bands.  We use a modified
self-consistent scheme, referred to as quasiparticle$-GW$
(QP$-GW$)-scheme in which $G$ and $W$ are calculated from an
approximate self-energy $\Sigma^{t}_0(\omega)=US\tilde{\tau_1}+
\left(1-Z^{-1}\right)\omega\tilde{\bf{1}}$, where the
renormalization factor $Z$ is adjusted self-consistently to match
the self-energy $\Sigma^{t}$ at low
energy.\cite{markiecharge,markiewater,arpes}  The vertex
correction $\Gamma({\bf k},{\bf q},\omega,\omega_p)$ in Eq.
\ref{selfeng} is taken as its first order approximation (Ward's
identity) as $\Gamma({\bf k},{\bf q},\omega,\omega_p)=1/Z$. Since
the $k$-dependence of $\Sigma$ is weak\cite{markiewater}, we
further simplify the calculation by assuming a $k$-independent
$\Sigma$, which we calculate at a representative point
$k=(\pi/2,\pi/2)$.

Our model is nearly parameter free.  We take the dispersions
directly from the LDA calculations, accurately fitted by a one
band tight-binding model\cite{markietb}, without any adjustment of
the resulting parameters. The susceptibilities are calculated
numerically, with no further approximations, while the
renormalization constant $Z$ is determined self-consistently.  We
have performed realistic Kanamori screening calculations due to
charge fluctuations to obtain a self-consistent value of $U$ and
order parameter $\phi$ at each doping starting from a doping
independent bare $U=1.7$~eV which reproduces the charge transfer
gap at half-filling.\cite{kanamori,markiecharge}

The optical conductivity, $\sigma (\omega)$, is calculated using
the standard linear response theory using the above
$\Sigma-$dressed single particle states. At high energies, we
subtract off a linear-in-$\omega$ background from the experimental
spectra associated with interband transitions to higher-lying
bands not included in the present one-band calculations,
consistent with supplementary Fig.~4 of Ref. \onlinecite{comanac}.
While keeping fixed the calculated weights of the Drude peaks, we
have phenomenologically broadened the spectra with additional
constant impurity scattering rates, similar to the universal form
found in experiments on NCCO\cite{onoseprb}, LSCO\cite{uchida},
and Bi2212\cite{hwang}.

\section{Discussion and Conclusion}

While the $(\pi,\pi)$ AFM order is known to be a robust feature in
electron doped cuprates,  in the hole doped case the exact nature
of the pseudogap is unknown, and is likely to be incommensurate
and possibly associated with charge order.  This is consistent
with our earlier study of ARPES\cite{tanmoytwogap}, where we
compared several possible competing ordered phases and
demonstrated that the shape and doping dependence of the pseudogap
is remarkably insensitive to the particular order so long as the
competing order vanishes in a quantum critical point near optimal
doping.  This conclusion should hold even more strongly for the
optical spectrum.  Note that we have included all the charge and
spin fluctuations in our calculations -- the particular leading
instability will only lead to a small rearrangement of spectral
weight in the low energy regime, leaving most of the optical
spectrum virtually unchanged.

We note that recent Gutzwiller approximation (GA) + RPA
calculations find a number of nesting instabilities in hole-doped
cuprates, resulting in a number of competing, incommensurate
density-wave orders.\cite{GARPA} Since the instabilities depend on
the bare susceptibilities, they are located at virtually the same
$q$-values in both the spin and charge (electron-phonon) channels.
Thus, the resulting optical spectra should be very similar in both
cases.

Finally, we note that the same model can be used to explore the
role of magnetic fluctuations as the pairing bosons for
superconductivity.\cite{markiesc} While the study of Ref.
\onlinecite{markiesc} revealed a significant contribution from
magnetic fluctuations, it was found that fluctuations in different
frequency ranges could act cooperatively to greatly enhance $T_c$.
This suggests that phonons could play a similar cooperative role
which could explain the isotope effect.\cite{lee}

In summary, we have shown that our model framework explains the
salient features of the optical spectra of cuprates and their
evolution with doping at a quantitative level. The rapid loss of
high energy spectral weight and the associated shift of the MIR
peak to low energies in the optical spectra with increasing doping
reflects collapse of the pseudogap order [here taken as a $(\pi
,\pi )$ AFM], and the presence of a QCP near optimal doping in the
coherent in-gap states. By contrast, the Mott gap in the
incoherent states persists at all dopings including, in
particular, the overdoped regime. The aforementioned coherent and
incoherent states are connected via a high-energy kink driven
predominately by magnetic excitations. In the magnetic excitation
spectrum, the dominant excitations lie in the waterfall region,
while the remnant of the low-energy magnetic resonance
mode\cite{lee} is less significant\cite{markiewater}.  Our model
self-energy scheme would provide a tangible basis for modelling
other spectroscopies (e.g., angle-resolved
photoemission\cite{arpes,waterfall}, scanning-tunnelling\cite{tanmoytwogap},
inelastic light scattering\cite{RIXS},
positron-annihilation\cite{positron}) of the cuprates and other
materials.

\begin{acknowledgments}
This work was supported by the US Department of Energy, Basic
Energy Sciences contract DE-FG02-07ER46352, and benefited from the
allocation of supercomputer time at NERSC and Northeastern
University's Advanced Scientific Computation Center (ASCC). RSM's
work has been partially funded by the Marie Curie Grant
PIIF-GA-2008-220790 SOQCS.
\end{acknowledgments}

\end{document}